# Nanoscale *β*-Nuclear Magnetic Resonance Depth Imaging of Topological Insulators

D. Koumoulis,[1]  G.D. Morris,[2] L. He,[3] X. Kou,[3] D. King Jr,[1] D. Wang,[7] M.D. Hossain,[2,7]

K.L. Wang,[3] G.A. Fiete,[4] M. G. Kanatzidis,[5] L.-S. Bouchard[1,6,*]

[1]Department of Chemistry and Biochemistry, University of California, Los Angeles, CA 90095, USA.

[2]TRIUMF, 4004 Wesbrook Mall, Vancouver, BC V6T 2A3, Canada.

[3]Department of Electrical Engineering, University of California, Los Angeles, CA 90095, USA.

[4]Department of Physics, University of Texas at Austin, Austin, TX 78712, USA.

[5]Department of Chemistry, Northwestern University, 2145 Sheridan Rd., Evanston, IL 60208, USA.

[6]California NanoSystems Institute, UCLA.

[7]Department of Physics & Astronomy, University of British Columbia, 6224 Agricultural Road, Vancouver, BC V6T 1Z1, Canada.

*Correspondence to:  Louis Bouchard (louis.bouchard@gmail.com)

Last revision: May 18, 2015

ABSTRACT

**Considerable evidence suggests that variations in the properties of topological insulators (TIs) at the nanoscale and at interfaces can strongly affect the physics of topological materials. Therefore, a detailed understanding of surface states and interface coupling is crucial to the search for and applications of new topological phases of matter. Currently, no methods can provide depth profiling near surfaces or at interfaces of topologically inequivalent materials. Such a method could advance the study of interactions. Herein we present a non-invasive depth-profiling technique based on $\beta$-NMR spectroscopy of radioactive $^8Li^+$ ions that can provide "one-dimensional imaging" in films of fixed thickness and generates nanoscale views of the electronic wavefunctions and magnetic order at topological surfaces and interfaces. By mapping the $^8Li$ nuclear resonance near the surface and 10 nm deep into the bulk of pure and Cr-doped bismuth antimony telluride films, we provide signatures related to the TI properties and their topological non-trivial characteristics that affect the electron-nuclear hyperfine field, the metallic shift and magnetic order. These nanoscale variations in $\beta$-NMR parameters reflect the unconventional properties of the topological materials under study, and understanding the role of heterogeneities is expected to lead to the discovery of novel phenomena involving quantum materials.**

# SIGNIFICANCE STATEMENT

The surface states of topological insulators (TIs) and magnetically doped TIs exhibit considerable inhomogeneities at the nanoscale. Methods are needed to probe the degree of heterogeneity as function of depth in nanoscale layers. We present a method that can directly visualize TIs in a depth-resolved manner and report on their electronic and magnetic properties. For example, in epitaxial thin films we demonstrate an increase in the density of states, a weakening of the ferromagnetic order when approaching the TI edges, as detected by measurements of the electron-nuclear hyperfine interaction, the effective $s$–$d$ exchange integral and local moment density. Depth profiling is expected to help uncover exotic physics of pure and ferromagnetic TIs and TI heterostructures.

## INTRODUCTION

Topological insulators (TIs) are narrow-gap semiconductor materials that are insulating in the bulk and conductive on their surface. The constituent atoms are typically heavy elements with large spin-orbit coupling (SOC). In time-reversal-invariant materials, the electronic structure of TIs is characterized by band inversions from strong SOC at an odd number of time-reversal invariant momenta in the bulk Brillouin zone. Unlike metals or ordinary insulators (OIs), charge carriers in TIs evolve from metallic to insulating as function of depth from the surface. The surface-state electrons are characterized by a suppression of backscattering and an intrinsic chirality of spin-momentum locking, making them of interest in spintronics. A number of theoretical predictions of experimental observations have been made including Majorana fermions, condensed-matter axions and quantized Hall conductance (*1,2*). Heterostructure engineering, where a crystal is formed consisting of a sequence of different building blocks (for example, alternating TI and OI layers), can trigger new physical phenomena such as TIs with enhanced bulk-band gaps (*3*) or Weyl semi-metals (*4*).

Because topological materials are characterized by sharp changes in electronic properties at their surfaces and at interfaces with other materials, sensitive techniques are required to probe electronic and magnetic properties in a spatially resolved manner down to the atomic length scale. It has become clear that TI properties are spatially dependent (both in-plane and as a function of depth or film thickness). Magnetic phenomena are accompanied by inhomogeneities within the material and interactions at interfaces that cannot be explained by simple models. To date, the observation of metallic surface states has been accomplished with transport measurements, scanning tunneling microscopy (STM), and angle-resolved photoemission spectroscopy (ARPES) (*5,6*). The depiction of the Dirac dispersion with ARPES requires *n*-type samples. Likewise, transport and STM require sufficiently conductive samples. Moreover, such experiments require the growth of high-quality samples (ideally, thin films or cleaved crystals exposing atomically flat surfaces) and often are most effective at low temperatures (<30 K) and for thin (<15 nm) layers. In addition, these techniques do not provide depth-resolved information. These factors limit their applicability in the study of complex heterostructures. Consequently, spatially resolved, non-invasive measurements of material properties would be an important asset in the study of physics at interfaces.

Nuclear magnetic resonance (NMR) has the potential to overcome some of the challenges associated with traditional characterization methods. Because it acts as a local probe of electron-nuclear hyperfine interaction with electrons near the Fermi level, it can depict the electronic and magnetic properties of insulating or metallic materials (*p*-type or *n*-type), does not require long-range crystal order and does not rely on electron transport. NMR has been regarded as an unlikely candidate for the study of topological states because of the low dimensionality of thin films and associated sensitivity issues: the detection of solid-state NMR signals typically requires at least $10^{15}$ nuclear spins whereas the nanometer-thick layers associated with topological surface states contain significantly fewer spins. A potential solution to the sensitivity problem is the technique of *β*-detected NMR (*β*-NMR) (see *SI Text*, *β-NMR experiments* and refs. *2-7*). *β*-NMR enables us to interrogate the wavefunction of the charge carriers while varying the energy of incident ions to control implantation depth. *β*-NMR is similar to muon-spin rotation spectroscopy, except that a $^{8}Li^{+}$ ion is used rather than a muon. The heavier mass of $^{8}Li^{+}$ enables better control of the ion position by controlling the energy of the incident ions. The implantation profile of the ions in high-$Z$ materials is sharp and localized to layers that are tens of nanometers thick. The signal from such thin layers can be detected with $\beta$-NMR using highly polarized nuclear spins (>60 %) and the high-efficiency detection of $\beta$-emissions. Furthermore, the longer half-life of the [8]Li isotope (838 ms) can reveals dynamics of magnetic field fluctuations or spin precessions over timescales longer than muons (half life, 2.2 µs) would allow. The only NMR studies of TI surface states published to date were done on nanocrystals (*7*) and nanowires (see *SI Text*, ref. *16*), by inferring the TI properties in the limit of high surface-to-volume ratios. A recent study on $Bi_2Se_3$ performed at high magnetic fields provides the means of approaching the quantum limit in TIs via NMR (see *SI Text*, ref. *17*). Because films grown by molecular beam epitaxy (MBE) and cleaved surfaces from single crystals are universally studied by most researchers, NMR experiments that directly interrogate TI properties in a depth-resolved manner beneath the surface of epitaxial TI films would be preferred, as they would enable comparison with the literature. Non-invasive NMR measurements of TI properties reflect intrinsic material properties and could even help in the study and control of spin polarized electron states. Herein, we investigate the possibility of depth-resolved measurements of both electronic and magnetic properties of epitaxial TI films using the technique of $\beta$-detected NMR ($\beta$-NMR) (see *SI Text,* refs. *2-7*). This technique may enable sensitive, non-invasive studies of surfaces and interfaces in topological phases. Additional advantages of the $\beta$-NMR technique may include less stringent requirements in terms of carrier type and concentration, operating temperature (*T*), film thickness, and film quality, compared with existing tools.

**EXPERIMENTAL METHODS**

Epitaxial thin films of the TI bismuth antimony telluride, $(Bi,Sb)_2Te_3$, and Cr-doped (~8%) $(Bi,Sb)_2Te_3$ (denoted CrTI below) with Bi/Sb ratios of 0.51/0.49 and 0.54/0.38 respectively, were grown on GaAs substrates (Fig. 1A,B). The Bi-to-Sb ratio and the Cr doping level were deliberately chosen so that Fermi level positions of the as-grown samples are already close to the Dirac point (see *SI Text,* refs. *19, 20*). Accordingly, we were able to demonstrate and realize electrical conduction dominated by spin-polarized surface states (see *SI Text,* refs. *21*), quantum interference competition (see *SI Text,* ref. *22*), quantum oscillation (see *SI Text,* ref. *19*), quantum Hall effect (see *SI Text*, section **Magneto-transport measurements**), and quantum anomalous Hall effect (see *SI Text,* ref. *22*) in the quantum limit regime. In this study, GaAs was chosen for two reasons. First, it is a suitable substrate for growth of this TI. Second, GaAs is a diamagnetic OI layer that provides an *in situ* reference for the $\beta$-NMR experiment, as probed by the beam energy of 19.9 keV. Its frequency shift as a function of depth and $T$ via $\beta$-NMR experiments is well studied and understood (*11*) (see *SI Text*). $\beta$-NMR measurements were conducted at the ISAC-I Facility in TRIUMF (Vancouver, Canada) using a beam of highly polarized radioactive $^8Li^+$ ions. Beam energies in the range 0.4–19.9 keV were selected to probe the film properties as a function of depth (Fig. 1A). In the present study, the beam energy of 0.4 keV was used to probe primarily the surface layer of the TI film (~3–5 nm implantation depth), whereas the 1-keV beam probed the bulk of the TI (ions implanted ~10–20 nm deep). Higher beam energies were used to probe deeper layers into the OI substrate. For additional details of the experimental procedure, see the *SI Text* (*Section A*).

## RESULTS AND DISCUSSION

**ELECTRONIC PROPERTIES**. The Knight shift is a NMR parameter that probes the local polarization of conduction band electronic spins induced by an external field. The nuclear spins are coupled to the conduction band electrons through the hyperfine interaction, which is a measure of carrier density (*8-10*). Quantitative values of the Knight shift, $K_d$, are obtained by correcting the NMR resonance frequency shift, $K = \frac{\nu - \nu_{ref}}{\nu_{ref}}$, for the demagnetization field, $\left(\frac{8\pi}{3}\right) \cdot \chi$, in the thin film (Fig. S2) according to

$$K_d = K + \left(\frac{8\pi}{3}\right) \cdot \chi, \tag{1}$$

where $\chi$ is the magnetic susceptibility. $K$ arises from coupling the $^8Li^+$ to the temperature (*T*)-independent Pauli susceptibility of the host nuclei, relative to the GaAs *in situ* reference. Similar to GaAs, $(Bi,Sb)_2Te_3$ is diamagnetic and exhibits a weak *T*-dependent magnetic susceptibility. Using the value of $\chi$ for $(Bi,Sb)_2Te_3$ reported by Stepanov *et al.* (*12*) and Van Itterbeek *et al.* (*13*), we note that in the data from refs. (*12*) and (*13*) $\chi$ follows a diamagnetic behavior over the range 2–300 K. The values of the Knight shift are plotted in Fig. 2A (upper inset) relative to the GaAs reference (see *SI Text*, *β-NMR experiments*). An alternate graphical presentation of these results for the pure TI is also shown in Section B of *Supplementary Materials*. The salient feature of this result is the substantially larger (negative) Knight shift near the surface of the TI film compared to the bulk. Such increased metallic shifts when approaching the surface of a TI are consistent with results from a previous study of diamagnetic $Bi_2Te_3$ nanocrystals (*7*). Knight shift measurements on $Bi_{0.5}Sb_{1.5}Te_3$ nanocrystals with conventional NMR (see *Fig. S6A*) also confirm the emergence of a negative Knight shift (see *SI Text, β-NMR experiments*) at the exposed surface. In the case of a metal, the wavefunction of delocalized charge carriers interacts with the nuclear spins through the electron-nuclear hyperfine interaction. Therefore, nuclear spins experience the average field of the electronic spin polarization, which leads to the Knight shift. The expression for the Knight shift in a degenerate semiconductor is (*15, 22, 23*)

$$K = \frac{4\pi}{3} g . g^* . \mu_B^2 . \langle |u_k(0)|^2 \rangle_{E_0} \rho(E_F), \tag{2}$$

where $g, g^*$ are electron $g$-factors (*vide infra*), $\rho(E_F)$ is the density of states at the Fermi level, and $\langle |u_k(0)|^2 \rangle_{E_0}$ is the single-particle free electron probability density at the nucleus. The latter is averaged near the bottom of the conduction band for electrons or near the top of the valence band for holes (*22*) and depends on the origin of the hyperfine interaction (Fermi contact, dipolar or orbital) (*15*). In our case, the nuclei are high-$Z$ elements, and the relativistic effects on the hyperfine coupling are of major importance (*15*). Therefore, the term $\langle |u_k(0)|^2 \rangle_{E_0}$ should be relativistically expressed as $(\cos^2 \theta^+) \langle R | \Delta(\vec{r}) | R \rangle$, where $\cos \theta^+$ is a spin-orbit mixing parameter, $R$ is the spatial atomic wave functions near the nucleus and $\Delta(\vec{r})$ replaces the Dirac function $\delta(\vec{r})$. The typical Knight shift is proportional to the paramagnetic susceptibility, although this proportionality is valid only for a scalar $g$ matrix (*15*). The spin-orbit parameter mixes the electronic wavefunctions thus mixing the hyperfine contributions that generate the Knight shift. In the non-relativistic approximation the hyperfine Hamiltonians use the free-electron $g$ factor rather than the modified $g^*$ because of the spin-orbit interaction and the crystalline potential (*15*). The magnitude and sign of the Knight shift in a narrow gap semiconductor are governed by the carrier density and the (large) $g^*$ factors of the carrier types, respectively. Equation (2) underestimates the Knight shift because it neglects demagnetization field effects near the TI surface and the non-parabolicity of the multi-valley band structure, which modulates the dependence between the Knight shift and the carrier density (*15,22,23*).

The electron-nuclear hyperfine coupling constant $A_{hf}$ and the Knight shift $K$ are related by

$$A_{hf} = \frac{K \cdot N_A \cdot \mu_e}{n \cdot \chi},$$ (3)

where $N_A$ is Avogadro's number, $n$ is the average coordination number at the implantation site, and $\mu_e$ is the magnetic moment of the charge carriers (see *SI Text,* refs. *4* and *14*). We use the value $n$=2.4 reported for lattice sites in $Bi_2Te_3$ (*14*). The hyperfine coupling, which is a local probe of the electronic wavefunctions in the vicinity of implanted $^8Li^+$ ions, is mediated by the strong SOC between the nuclei and the *p*-band carriers (*8,15-17*). The hyperfine constants are plotted in Fig. 2A. We observe a *T*-dependent $A_{hf}$, which approximately doubles from the TI bulk (1 keV) to the TI surface (0.4 keV) that holds over the entire temperature range. This doubling indicates a higher (2×) carrier concentration at the TI surface compared with the TI bulk. We note that the hyperfine coupling constant of $^8Li^+$ measured near the surface of the TI approaches the value measured in a thin metallic silver film (see *SI Text*, ref. *14*), 20.5 kG/$\mu_B$ (octahedral site) at low *T*. These results are consistent with transient reflectivity studies of $Bi_2Se_3$ thin films with varying thicknesses. These previous studies revealed a similar insulator (bulk TI) to metal (surface TI) crossover with the film thickness decreasing from 25 nm to 6 nm.  In addition, the ultrafast carrier dynamics at the surface (~6 nm) reached values comparable to those observed in noble metals (*18*).

Similarly, the behavior of a trivial insulator under *β*-NMR is qualitatively different from that of the TI.  In contrast to the observed increased metallic shift when approaching the surface of a TI, *β*-NMR studies of MgO (a trivial insulator) thin films yielded no evidence of depth-dependence of its resonance frequency as function of beam energy (see *Fig. S5*, and *SI Text*, Section B).  Other detailed NMR studies of the Knight shift and relaxation in PbTe ($E_g$=0.32 eV, 300 K) and Tl$_2$Se ($E_g$=0.6 eV, 300 K) were consistent with a decrease in carrier concentration in nanoscale materials compared with bulk materials (*20,21*), in contrast to the behavior observed in TI nanocrystals. Furthermore, in the case of ZnTe ($E_g$=2.23 eV, 300 K) the [125]Te frequency shift of *nano*-ZnTe remains unshifted relative to micrometer-sized samples (see *Fig. S6C*).  Finally, in a *β*-NMR investigation in GaAs ($E_g$=1.4 eV, 300 K) in which the effect of implantation energy was studied, no detectable Knight shift was reported in the case of *n*-GaAs, as the beam energy decreased from 28 keV to 3 keV (*see SI Text,* ref. *5*).  In that same study, no concomitant line broadening was observed, as would be expected of a transition to a metallic state.  The lack of line broadening is in clear contrast to the TI layer depth-resolved properties measured in this study.  This collection of experiments on trivial insulator surfaces and nanocrystals clearly indicates distinctly different and non-metallic behavior compared with the phenomena observed at the surface of the TI.

The estimates of the Knight shift by equation (2) (Fig. 2 inset) are reasonably close to values previously reported (*24–26*) in transport studies, namely, ~10^{19} cm^{-3} for Bi$_2$Te$_3$ and ~10^{18} cm^{-3} for (Bi,Sb)$_2$Te$_3$ thin films.  Surface carrier concentrations were reported to be in the range $10^{11}$-$10^{12}$ cm$^{-2}$ (*24–26*), of the same order as observed here. Higher carrier concentrations at higher temperatures (150–288 K) have been commonly observed in narrow-gap semiconductors and are attributed to a crossover from a thermally activated regime to a saturation regime.

A plot of the linewidth as a function of *T* and beam energy is shown in *Figs. S2 & S3B*. In the GaAs layer, the dependence of the linewidth on temperature is considerably small, similar to results from previous *β*-NMR studies in intrinsic GaAs (*11*). However, in the TI layer (0.4-1 keV), the linewidth is more than three times greater than in GaAs. We also note that the linewidth in the TI layer increases with decreasing *T*, in contrast to the non-TI material, GaAs (*11*). Another feature we observe is a considerably broader linewidth for the TI surface layer compared to the TI bulk, a result that is opposite of that observed for trivial (non-TI) semiconductors (see *SI Text*, ref. *4*).

**Magnetic Properties**. We now turn our attention to the magnetic properties and examine the case of the Cr-doped TI thin film. The ability of *β*-NMR to differentiate pure TI from magnetic TI is clearly demonstrated by the *T*-dependence of the Knight shift (Fig. 3A) and linewidth (Fig. 3C), which is substantially different for these two films, from 20 K up to ambient *T*. This trend is compatible with the development of magnetic correlations, which grow progressively stronger as *T* is lowered.

The second observation is that in the undoped TI, signals near the surface exhibit a larger negative Knight shift than in the TI bulk whereas with Cr doping, the opposite is true; no increase in Knight shift is apparent at the surface (see *Figs. S2 and S4A*), consistent with the behavior of gapped surface states. In the Cr-doped TI, the Knight shift is promoted by the 3*d* spins (transferred field), diamagnetic core contributions as well as *s*-like contributions from conduction electrons. The [8]Li *β*-NMR frequency as a function of temperature describes the influence of the Cr dopants. We observe an inflection point near 75 K for the surface layer and 150 K for the bulk-like layer that is due to the loss of magnetic order. Previous studies via magnetic and transport characterizations have predicted a robust ferromagnetism close to room temperature for Cr-Bi$_2$Te$_3$, approximately 100 K for Bi$_2$Se$_3$:Mn and up to 190 K for Sb$_{2-x}$Cr$_x$Te$_3$ (*27-30,34*). In non-zero external magnetic fields, the loss of magnetic order tends to be more extensive and shifts to higher temperatures compared with the case of a zero external field (*28-33*) owing to the creation of large local fields both above and below the critical temperature. Although no proper phase transition occurs in non-zero external fields in terms of critical phenomena, demagnetization nonetheless occurs, and we write "$T_c$" to refer to the region of inflection.

We now discuss the line broadening. In the *β*-NMR experiment, the presence of magnetic moments at the dopant sites will generate local internal fields that broaden the [8]Li *β*-NMR resonance. This broadening is proportional to the magnitude of the local magnetic moment density and reflects the distribution of this field. Whereas the frequency shift reflects the average field within the measurement region, the linewidth reflects the *rms* field fluctuation. To exclude nonmagnetic sources of line broadening such as a quadrupolar broadening (since [8]Li$^+$ has a small quadrupole moment), power broadening etc., we perform the analysis of the fractional broadening as defined by the relative linewidth parameter ($\Delta w = \frac{W_{CrTI} - W_{TI}}{W_{TI}}$), where $W_{TI}$ is the linewidth of the [8]Li resonance in the TI film and $W_{CrTI}$ is its linewidth in the Cr-doped TI film.

The results for $\Delta w$ are shown in Fig. 3B. For the surface layer, the temperature dependence of linewidth, in accordance with the resonance shift, unveils the onset of a disparity near 75 K (Fig. 3B). Notably, the $^8$Li $\beta$-NMR line broadens in the surface and bulk regions by approaching the transition temperature from below. This important observation may be associated with the presence of a small amount of local ordering that causes appreciable dipolar broadening above $T_c$. However, the possible presence of lattice distortions and strain effects in thin films cannot be completely excluded. At 1 keV energies (in the CrTI bulk), $\Delta w$ suggests that the magnetic transition $T_c$ occurs near 150 K, approximately two times higher. This result is consistent with the prediction of molecular-field theory that $T_c$ should be proportional to the number of nearest neighbors, which is two times larger in the bulk than on the surface (*31*). Voigt and co-workers (*35-38*), using differential perturbed angular correlation measurements of the magnetic hyperfine field in the topmost Ni monolayer and that in the deeper layers, also reported a similar increase in $T_c$. Similarly, we observed the transition temperature to decrease with decreasing penetration depth. This decrease in temperature could be due to the reduced disorder-causing fluctuations and smaller coordination number at the surface relative to the bulk.

The *T*-dependence of $\Delta w$ yields an estimate of the magnitude of the effective *s–d* exchange integral (*J*) by the relation $\frac{W_{CrTI}-W_{TI}}{W_{TI}} = \left(\frac{Jp^2C}{3gk_B}\right) \cdot \frac{1}{T} + D$, where *p* is the effective number of Bohr magnetons, *C* the atomic fraction of paramagnetic atoms, *D* a temperature-independent term, $k_B$ is the Boltzmann constant, and *g* is the *g* -factor *(36, 37)*. This relationship provides the value of *J* that emerges from the Cr localized moments in the $(Bi,Sb)_2Te_3$ matrix. The slopes $\left(\frac{Jp^2C}{3gk_B}\right)$ were found to vary slightly from $\left(\frac{Jp^2C}{3gk_B}\right)^{bulk}$ =5.07±1.76 K for the bulk to $\left(\frac{Jp^2C}{3gk_B}\right)^{surface}$ = 4.53±1.05 K for the surface layer; specifically, we observed that $J^{surface} = 0.9 \cdot J^{bulk}$. The value of the $J$ ratios (0.9) suggests that not only the chromium density but also the carrier density should be considered when describing the line broadening on the surface and in the bulk layer. The Rudermann-Kittel-Kasuya effect is approximately the same order of magnitude as the nuclear dipolar interaction. Therefore, the present results do not rule out the possibility of a contribution from carrier-mediated ferromagnetism since both the hole/electron concentration and the chromium density drastically affect the $T_c$ values independently (*27–30*). The presence of Cr in chalcogenides likely functions as a donor and increases the charge carrier concentration; therefore, our results are consistent with carrier-mediated magnetism in the surface of the Cr-doped TI film. In the case of different dopants trends for $T_c$ as function of film thickness were reported by using different techniques. For comparison, we mention the results of Zhang and Willis based on measurements of bulk film magnetism through the magneto-optical Kerr effect (*35*), which were theoretically verified by Rausch and Nolting (*38*) in the molecular field approximation (Weiss mean field theory) of the Heisenberg model. Furthermore, a recent ferromagnetic resonance study of $Bi_2Se_3$:Mn has revealed a stronger ferromagnetic order in the bulk than on the surface, which was attributed to the bulk layer being more homogeneously doped than the surface layer (*34*), in agreement with our observation. This result is also consistent with the larger fractional broadening $\Delta w$ for the 1 keV beam (bulk of the TI) compared with the 0.4 keV beam (near TI surface), recalling that the fractional broadening is indicative of local moment density.

Compared with other techniques, our experimental procedure differs mainly in that rather than varying the film thickness, the reading is performed non-invasively, that is, as a function of depth, and in a film of fixed thickness. This experimental aspect is missing, for example, in ARPES and STM. Therefore, the $\beta$-NMR spectroscopy is an efficient and valuable tool for TIs whose topological properties depend strongly on film thickness or for TI layers that are part of a nanoscale device or heterostructure.

## CONCLUSIONS

This depth-dependent study of electronic and magnetic properties of TI epitaxial layers using implanted $^8Li^+$ ions reveals important details about nanoscale layers. Namely, transitioning from bulk to the surface of the TI across a distance shorter than 10 nm, the electron–nuclear hyperfine coupling constant $A_{hf}$ approximately doubles, reaching a value ~16 kG/$\mu_B$ at 20 K, which is comparable to certain pure metals, intermetallic and cuprate compounds. This $A_{hf}$ doubling is accompanied by a large negative Knight shift and line broadening, both of which phenomena approximately double from bulk to surface. The advantage of the $\beta$-NMR output was most effectively illustrated by non-invasively extracting the magnetic properties of the TI as function of depth. As a function of temperature, the linewidth of the TI departs from that of the CrTI at approximately 75 K (for the surface region) and 150 K (for the bulk region) because of the presence of Cr dopants, a result consistent with a gapped TI surface. The CrTI exhibited a 10-% decrease in the effective $s$–$d$ exchange integral when moving from the bulk to the surface, a decrease that is explained by carrier-mediated magnetism. This experimental approach could prove useful to understanding depth dependent interactions, proximity-induced phenomena across interfaces and heterostructures in TIs, crystalline insulators, topological superconductors and interaction-driven topological phases.

## MATERIALS AND METHODS

**Thin Film Heterostructures Growth**.  $(Bi_xSb)_2Te_3$ thin films were conducted in a Perkin Elmer MBE system under ultra-high vacuum conditions.  Intrinsic GaAs (111) wafers ($\rho > 10^6$ $\Omega \cdot$cm) were cleaned by a standard Radio Corporation of America (RCA) procedure before being transferred into the growth chamber.  GaAs substrates were annealed in the chamber under Se-protective environment at ~580°C for 30 min. During growth, Bi, Sb and Te cells were kept at 470, 395 and 320°C respectively, while the GaAs (111) substrate was kept at 200 °C (growth temperature).  After growth, 1.5 nm of Al was subsequently deposited *in situ* at 20 °C to protect the epilayer from unintentional doping in the ambient environment.  Al film was later naturally oxidized to form $Al_2O_3$ after the sample was taken out of the chamber and exposed to air.  After oxidation, the final thickness of the $Al_2O_3$ capping layer was approximately 3 nm. The samples were cut to size 8 mm × 8 mm × 0.35 mm and mounted on sapphire substrates to provide good thermal contact and enable alignment of the $^8Li^+$ beam.

**Magneto-transport measurements**.  Four-point Hall measurements were conducted using a Quantum Design physical property measurement system (PPMS) at the base temperature of 1.9 K.  This setup enables us to systematically adjust several experimental variables such as temperature, magnetic field, measurement frequency, external gate bias, etc.  Multiple lock-in-amplifiers and Keithley source meters were connected to the PPMS system, enabling compressive and high-sensitivity transport measurements.  The samples used for transport measurements were patterned as thin film Hall bar devices.

**ARPES**.  The electronic structure of the $(Bi,Sb)_2Te_3$ thin film was studied by ARPES. The measurements were performed at BL12.0.1 of the Advanced Light Source Division (Lawrence Berkeley Lab, Berkeley, CA).  Samples (without Al capping) for ARPES were further conditioned by mild annealing at $T$ = 200 °C in the experimental chamber for two hours.  All photoemission data were collected from the samples at 10 K.

**$\beta$-NMR**.  The $\beta$-NMR experiment was performed at the ISAC-I Facility (Isotope Separator and Accelerator) radioactive ion beam facility (TRIUMF, Vancouver, Canada) using a beam of highly spin-polarized radioactive $^8Li^+$ ions (lifetime $\tau$=1.2 s, $I$=2, gyromagnetic ratio $\gamma/2\pi$=6.3015 MHz/T).  The high magnetic field spectrometer (up to 9 T) is mounted on a high voltage platform allowing the application of a retarding electrostatic potential, slowing the incoming $^8Li^+$ ions before implantation into the sample.  Beam energy can be adjusted from 20 keV down to 0.33 keV for depth-resolved investigations.  Samples were mounted on a coldfinger cryostat situated at the center of the magnet and in vacuum of $10^{-10}$ mbar.  Nuclear polarization – initially anti- parallel to beam momentum – was monitored via the parity-violating $\beta$-decay of the $^8Li$ nucleus ($^8Li \rightarrow {}^8Be + e^- + \overline{\nu}_e$) in which the momenta of outgoing betas are correlated with the instantaneous orientation of nuclear spin.  Two beta-particle detectors situated upstream ($B$) and downstream ($F$) with respect to the sample recorded the decay events.  The experimental signal proportional to nuclear polarization is the asymmetry in decay events counted by the two beta detectors, $P \propto \frac{(n_B - n_F)}{(n_B + n_F)}$. $\beta$-NMR spectroscopy was carried out at a fixed magnetic field ($H_0$=6.55 T, applied normal to the sample face) by slowly stepping the radiofrequency magnetic field ($H_1$=35 $\mu$T parallel to the sample face) repeatedly up and down through the resonance condition at a rate of 2 kHz/s.  Each run accumulated approximately $5 \times 10^9$ events. Near resonance, incoherent transitions among the nuclear spin states driven by the continuous RF result in the destruction of nuclear polarization and a dip in asymmetry. The $\beta$-NMR lineshape reflects the magnetic field distribution probed by the $^8$Li within the thin film.

## ACKNOWLEDGMENTS

This article contains supporting information online. The research at UCLA and NU was funded by DARPA MESO, Award No. N66001-12-1-4034. The magnetometry measurements were funded by AFOSR and DARPA QuASAR. G.A.F. was funded under ARO W911NF-14-1-0579 and NSF DMR-0955778. L.S.B. acknowledges useful discussions with S.D. Mahanti. D.K. acknowledges W.A. MacFarlane, R. Kiefl for their help and guidance during the $\beta$-NMR experiments and for providing the MgO results ahead of publication as well as B. Leung and R. E. Taylor for their help during the NMR and PXRD studies on $Bi_2Te_3$ and $Sb_2Te_3$. All authors acknowledge the use of instruments at the Molecular Instrumentation Center (MIC) facility at UCLA. D.K acknowledges B.J. Archer for drawing the illustration of Fig. 1.

**FIGURE LEGENDS**

**Figure 1. Schematic diagram of the *β*-NMR experimental set-up**. The direction of the $^8$Li$^+$ ion beam and the different epitaxial layers are shown in (**A**). Structure of the TI-OI multilayered $(Bi,Sb)_2Te_3$ sample [capping layer (< 3nm $Al_2O_3$)], TI: topological insulator (50 nm $(Bi,Sb)_2Te_3$) and OI: ordinary insulator (350 μm GaAs) as a function of depth (nm) (**B**).

**Figure 2. Depth and temperature (*T*) dependence of the hyperfine coupling constant (*$A_{hf}$*).** Surface and bulk measurements were made at beam energies of the surface (0.4 keV) and bulk (1 keV), respectively. The upper inset shows the Knight shift (*K*) as a function of *T* for surface (red circles) and for 1 keV (black polygons) with respect to the resonance position of GaAs (19.9 keV, an *in situ* reference) (**A**). The lower inset shows estimates of the local carrier density. *β*-NMR response of the CrTI film compared with undoped TI as function of depth and inverse temperature (**B**). Error bars are smaller than the symbol sizes to all figures.

**Figure 3. Temperature dependence of Knight shift and relative linewidth at the surface of TI and CrTI.** (**A**) Temperature dependence of Knight shift at the surface of pure TI (open red stars) and Cr-doped TI (filled black stars) measured with *β*-NMR. Inset: *β*-NMR spectra at 20 K. (**B**) The temperature dependence of the relative linewidth parameter (*see main text*, Section **Magnetic properties**) of $^8$Li$^+$ for beam energies 0.4 keV (blue squares) and 1 keV (green circles). (**C**) The temperature dependence of the linewidth for surface of undoped (black squares) and Cr-doped TI (red circles) reflects the effect of Cr dopants below 75 K. Experimental uncertainties are smaller than the symbols for all displayed figures.

**FIGURES**

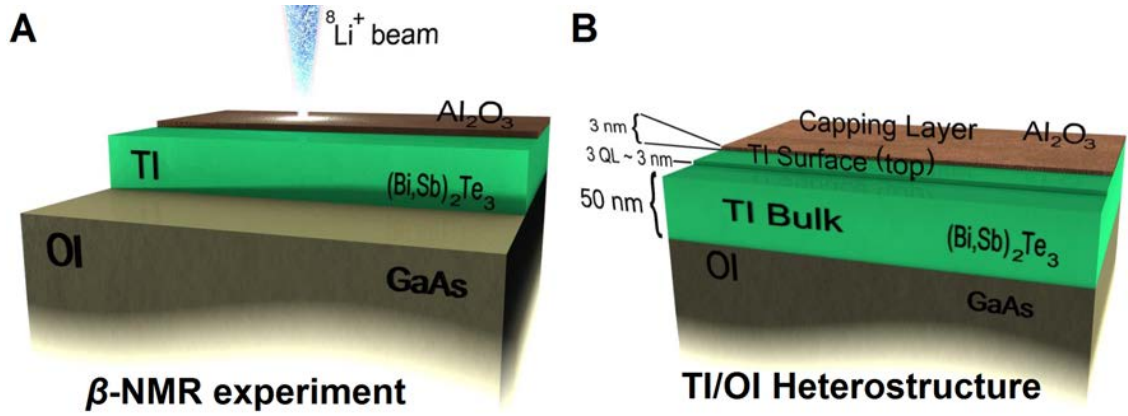

**Figure 1.**

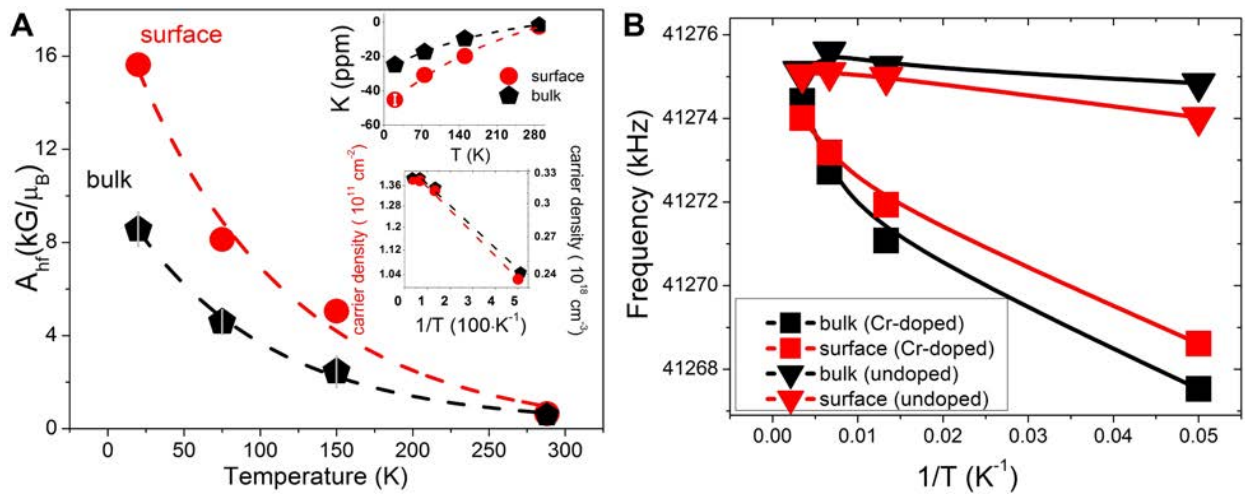

**Figure 2.**

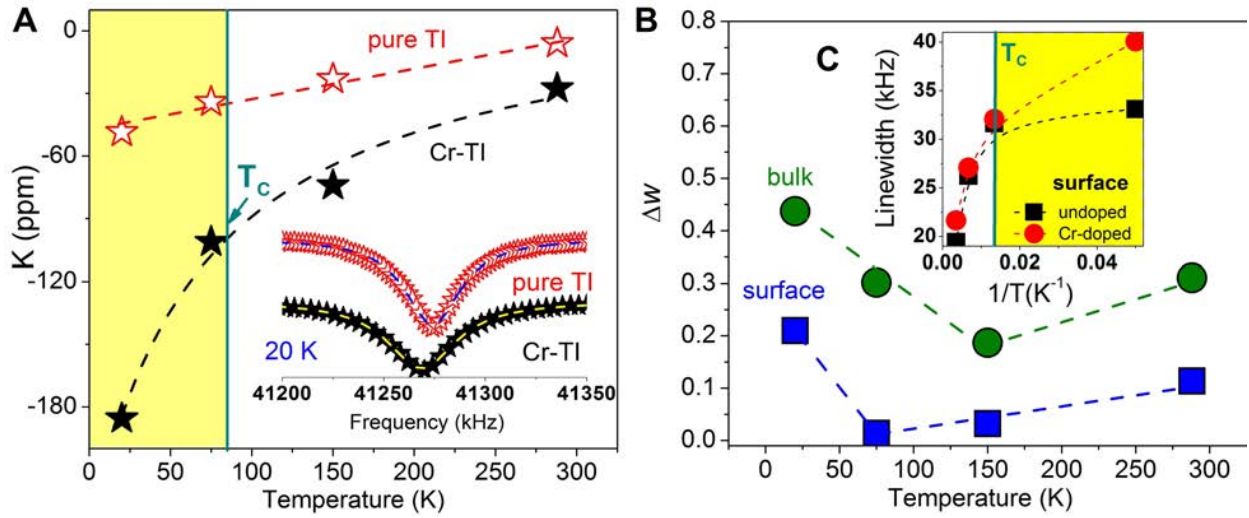

**Figure 3.**

**SUPPORTING INFORMATION**

**SI Text**

## A. Experimental Setup, Sample Growth and Characterization

**Samples**. Bismuth antimony telluride, $(Bi,Sb)_2Te_3$, has a rhombohedral structure with space group $D_{3d}^5$ ($R\overline{3}m$) described by perpendicular layers which form an ABC stacking structure with five atoms per unit cell. There are no symmetrical lattice sites and cubic positions in this crystal group. The quintuple layers [QL, (1QL~1nm)] forming the lattice feature two different crystallographic sites for the tellurium Fig. S1A. The ternary alloy $(Bi,Sb)_2Te_3$ is of considerable interest (bulk band gap ~0.3 eV) because it features a substantially improved insulating bulk (see *Main Text*, ref. *5*) compared to the parent binary compound, $Bi_2Te_3$. While the addition of Sb to bismuth telluride alters the Fermi level ($E_F$) position, the surface Dirac cone remains hexagonally warped (see *Main Text*, refs. *5,6*), similar to $Bi_2Te_3$. The quintuple-layer structure of $(Bi,Sb)_2Te_3$ thin films is evident from the high-resolution transmission electron microscopy (HRTEM) result of Fig. S1B (see ref. *33*).

The electronic structure of $(Bi,Sb)_2Te_3$ was studied by ARPES at 10 K. The Dirac cone, which confirms the existence of metallic surface states, is observed from the ARPES intensity map (Fig. S1C). The Fermi level, which lies inside the bulk conduction band, has been observed previously (*1*). Magneto-conductance transport measurements were carried out at 1.9 K, with external field applied perpendicular to the thin film, via four-point Hall measurements Fig. S1D. A sharp, weak anti-localization cusp appears in the low field limit, indicative of destructive quantum interference conduction along the topological surface at zero magnetic field.

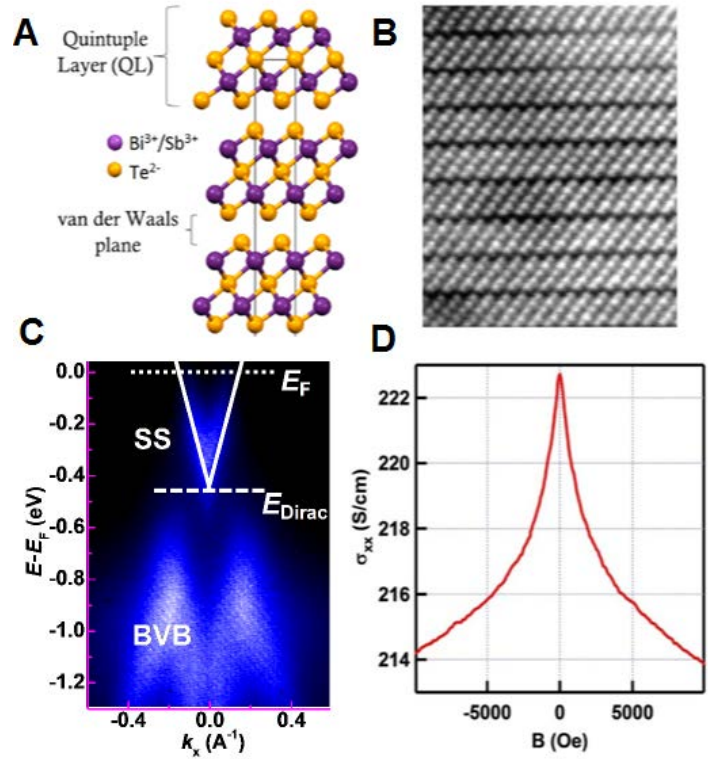

**Fig. S1. Characterization of the (Bi,Sb)$_2$Te$_3$ heterostructure.** (**A**) The hexagonal view of the unit cell of bismuth antimony telluride (Bi,Sb)$_2$Te$_3$, consists of perpendicular quintuple layers bound by weak van der Waals interactions. (**B**) Atomic resolution HRTEM micrograph depicts the quintuple-layered structure of the TI film (see also Ref. 33 for further characterizations). (**C**) The electronic structure of the (Bi,Sb)$_2$Te$_3$ thin film via ARPES. The Dirac cone can be clearly observed from the ARPES intensity map. The Fermi level is indicated by the horizontal dotted line. The position of the Dirac point is indicated by the intersection of the diagonal lines. Magneto-conductance measurements of (Bi,Sb)$_2$Te$_3$ thin films (6 QL) on GaAs substrate at 1.9 K. The external magnetic field is applied perpendicular to the samples. (**D**) A sharp negative weak anti-localization cusp appears under low field, indicating the destructive quantum interference conduction along the topological surface.

In this study, layered heterostructures (Fig. 1) were grown by MBE, which consist of the following sequence: **1.** 350 µm-thick intrinsic GaAs (111) OI layer, **2.** 50 nm-thick (Bi,Sb)$_2$Te$_3$

Depth-Resolved MRI of TIs

(or Cr-doped, ~8%) (Bi/Sb ratio of 0.49/0.51) TI layer and **3.** < 3 nm-thick $Al_2O_3$ capping layer (CL) to mitigate environmentally induced deterioration.

**$\beta$-NMR experiments**.  All samples for NMR were stored under vacuum and measured within three days of their preparation.  $\beta$-NMR experiments were carried out using a beam of highly polarized radioactive $^8Li^+$ ions.  The nuclear polarization was monitored by the anisotropic $\beta$-decay of the $^8Li$ nucleus (*2-7*).  The resulting $\beta$-NMR spectrum provides a direct measure of the electronic structure near and well beneath the sample's surface. The geometry of the experiment is illustrated in Fig. 1A.  Typical spectra of $^8Li^+$ for two different temperatures and beam energies are shown in Figs. S2A,B. All $\beta$-NMR spectra were well described by a Lorentzian function. Such Lorentzian lines in $\beta$-NMR experiments have been previously attributed to inhomogeneous broadening from magnetic field inhomogeneities, power broadening from the radiofrequency source, and dipolar interactions to magnetic host nuclei (*2-7*).

Beam energies were selected according to the Stopping and Range of Ions in Matter (SRIM) Monte Carlo predictions (*8*) of the $^8Li^+$ ion implantation depth. Five different beam energies (0.4, 1, 4, 10 and 19.9 keV) were selected to probe implantation depths down to 200 nm.  Because of the predicted mean implantation depth (4.2 nm) and ion straggle (3.1 nm), nuclear spin resonances from the lowest energy beam (0.4 keV) are expected to extract information about the TI and Cr-doped TI (CrTI) surface state, which is known to be located within the first 3 QL from the surface (see Main Text, Ref. *5*).  As the energy increases, the beam interrogates bulk TI (1 keV) and bulk OI layer ($\geq$10 keV).  About 50% of the $\beta$-NMR signal for the surface profile originates from the TI and the CrTI surface (3 QL), while the rest is from the TI bulk.  For the Depth-Resolved MRI of TIs 1 keV beam, more than 80% of the signal originates from the bulk TI (less than ~18% of the signal originates from the first 3 QL). SRIM predictions may underestimate the stopping range if channeling effects are important (*4*).

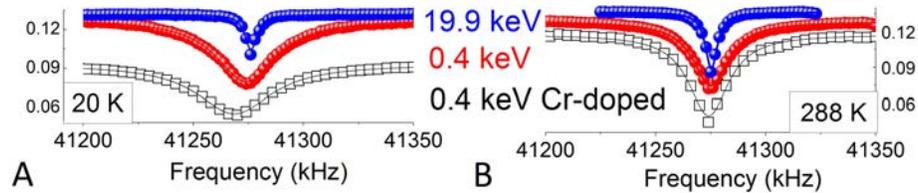

**Fig. S2. *β*-NMR investigation of the TI and CrTI heterostructures.** *β*-NMR spectra of pure TI, magnetically doped TI (surface) and OI (19.9 keV) at 20 K (**A**) and 288 K (**B**). *β*-NMR asymmetries at all temperatures and energies are best described by a single Lorentzian line. The vertical scale (*β*-NMR asymmetry) pertains to the red curve (pure TI). The remaining two spectra (Cr-doped TI and OI) are offset vertically for display purposes, so as to avoid overlap.

Figure S3 below presents results from the pure TI in the form of a color map. The stars indicate the position of our experimental measurements, whereas the colormap interpolates smoothly between these points for visualization purposes. The color scales indicate frequency (MHz) (Fig. S3A) and linewidth (kHz) (Fig. S3B).

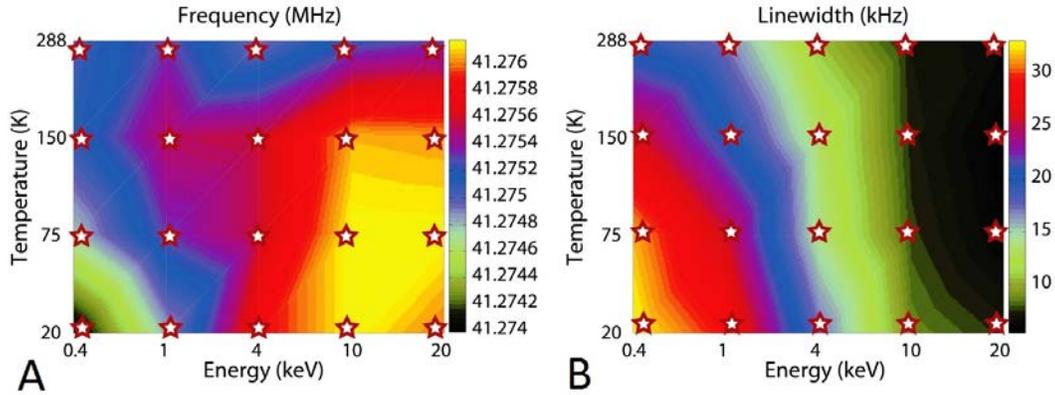

**Fig. S3. *β*-NMR measurements in pure TI film.** (**A**) Frequency and (**B**) Linewidth as a function of temperature and implantation energies. The TI bulk is best probed by the 1 keV beam, whereas the TI surface is reflected in the 0.4 keV beam data. The GaAs layer is probed by the 10 and 19.9 keV beams.

Plots of center frequency versus temperature and beam energy (line shapes & color maps) are shown in Fig. S4. Two main features can be observed: **1.** the NMR shift becomes more negative with decreasing beam energy, **2.** this NMR shift is universal for all temperatures we measured from 20 K to 288 K. Furthermore, this temperature dependence of the NMR shift with beam energy is more pronounced at low temperatures. In the intrinsic GaAs layer (19.9 keV beam energy), the carrier concentration is exceedingly low ($\sim 10^7$ cm$^{-3}$). As a result of the low carrier concentration, there is no Knight shift and the frequency is independent of temperature below 150 K (*4*). The shift in GaAs increases slightly above 150 K, in agreement with previous results (*5*), due to a site change for $^8$Li$^+$ ions from an interstitial to a substitutional site above 150 K. Such negative Knight shifts in narrow-band multi-valley semiconductors have been explained previously in terms of spin-orbital coupling (*9,10*). Moreover, a recent $^{71}$Ga MAS-NMR study of *nano*-GaN (band gap $E_g$=3.4 eV at 300 K) revealed no *T*-dependent frequency shift, in contrast to the $^{71}$Ga Knight shift measured in gallium metal. If interactions of nuclear spins with conduction electrons were to play a dominant role, a *T*-dependence of the $^{71}$Ga NMR spectrum would have been expected, yet in experiments, the results were independent of *T* (see *Main Text*, ref. *19*). One of the defining features of such diamagnetic Knight shifts is their temperature dependence, which gravitates toward more negative frequencies as temperature decreases, owing to the temperature dependence of the energy gap (*9*). At the lowest beam energy (0.4 keV), which was chosen to probe the TI and CrTI surfaces, the diamagnetic shift is largest. As pointed out by Misra and co-workers (*9*) the larger the carrier concentration, the more negative is the Knight shift. The increased diamagnetic shift at low energies, which scales with carrier concentration, is consistent with the higher carrier concentrations that are expected from TI surfaces (see *Main Text*, ref. *5*).

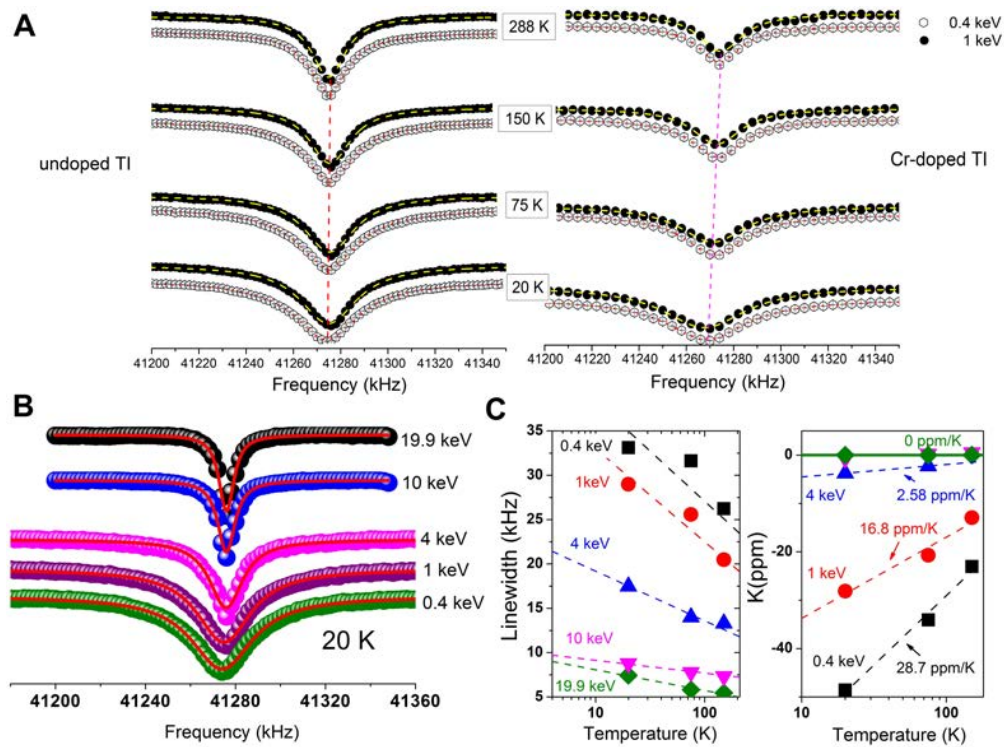

**Fig. S4.** *β*-NMR measurements in pure and CrTI films. (**A**) Lineshapes of pure and CrTI at 0.4 keV and 1 keV. (**B**) The depth profile of lineshapes of pure at 20 K and (**C**) Linewidth and Knight shift as a function of temperature and implantation energies. The GaAs layer is probed by the 10 keV and 19.9 keV beams.

The behavior of Cr-doped TI is significantly different, as seen in Figs. S2 and S4A. In the undoped TI, signal near the surface (0.4 keV beam) exhibits a larger negative Knight shift than the TI bulk (1 keV beam). With Cr doping, the opposite is true – the larger Knight shift at the surface is no longer apparent. For the undoped TI, the carrier density at 1.9 K is $8 \cdot 10^{11}$ cm$^{-2}$. For the Cr-doped TI, this number will be affected by the Cr doping concentration. For Cr (%) from 2 % to 20 %, the carrier density is known to increase from $0.1 \cdot 10^{13}$ to $2.2 \cdot 10^{13}$ cm$^{-2}$. For the present Cr concentration (8 %), the carrier density is in the range $3 \cdot 10^{12}$ to $5 \cdot 10^{12}$ cm$^{-2}$. This large increase in the carrier concentration explains the observed overall shift. While Cr dopants are known to create ferromagnetic order in TIs, the mechanism of interplay between the carrier (hole) density and phase transition temperature [carrier-mediated ferromagnetism in dilute magnetic semiconductor (DMS) or a carrier-independent ferromagnetism] remains an active area of research (see *Main Text*, refs. *24-27*). The insertion of Cr dopants into the chalcogenide matrix could also promote the formation of a dilute magnetic alloy (DMA), an effect strongly dependent on the Cr solubility limit in the host lattice. However, according to most recent studies of the addition of transition metal ions to the TI matrix, Cr is a uniformly distributed and thermodynamically stable dopant (see *Main Text*, ref. *27*) leading to long-range magnetic ordering (see *Main Text*, refs. *24-27*). The above statement is confirmed at the local level by a homogeneous broadening of the NMR line (Lorentzian shape), thus reflecting the local distribution of magnetic moments. The presence of phase separation (PS) or a DMA will generate different hyperfine couplings, thus leading to a PS and an inhomogeneous broadening of the NMR lineshape. Our observed broadening excludes the case of DMAs and PS in the Cr-doped TI film. Interestingly, despite the presence of an *n*-type semiconductor (GaAs) adjacent to a *p*-type TI material the *β*-NMR was not capable, in the present study, to unveil a conspicuous feature related to the TI/OI interface behavior. This inability can be attributed to the limited number of data points in this region of the heterostructure as the initial aim of our work was restricted to the vacuum-TI surface region. Further experiments utilizing more beam energies would be warranted to study this TI/OI interface.

## B.  *β*-NMR of [8]Li$^+$ Implanted in a Normal Insulator: The Case of MgO

The high field [8]Li$^+$ *β*-NMR resonance in rocksalt MgO is very narrow, with minimal nuclear dipolar broadening and no quadrupolar splitting, the latter implying a site with cubic symmetry for the implanted [8]Li$^+$. The resonance in MgO is commonly used as a frequency reference, from which the resonance shifts in other materials are measured. We present measurements of the high field lineshape and its energy dependence.

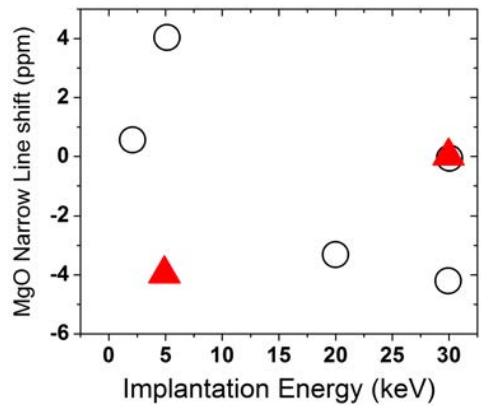

**Fig. S5.** The shift of the narrow MgO resonance as a function of [8]Li$^+$ implantation energy at 300 K and 3 T (open circles). The zero of shift is arbitrarily chosen to be zero for one of the 30 keV spectra at each field. Statistical error bars from the fits are included, but systematic uncertainty dominates and is indicated by the difference between the two 300 K values. Red triangles: at 4.1 T and 250 K in a separate experiment.

As the ion implantation energy is reduced, the implantation profile moves closer to the free surface. The dependence of the $^8$Li$^+$ $\beta$-NMR on implantation energy has been studied briefly on a few occasions, the most detailed at 300 K and 3.0 T. The low power (240 mW) spectra are well fit to the sum of a narrow and a broad Lorentzian. The broad resonance is slightly positively shifted, but not clearly resolved, and primarily accounts for the wings of the line. The low RF power de-emphasizes any broad component, due to the background signal (related to $^8$Li$^+$ backscattering that we expect below ~5 keV). The position of the narrow line is shown as function of implantation energy in Fig. S5. A second measurement was done at 4.1 T, 250 K and 900 mW. The results are shown as red triangles in Fig. S5. No evidence for depth dependence of the resonance position could be found. For further details, see *(18)*.

## C. Nanocrystalline Topological Insulators and Ordinary Insulators

In this section, we present variable temperature solid-state NMR experiments to corroborate the above findings from our $\beta$-NMR experiments. In nanoparticles of TI material, surface effects begin to dominate and the NMR signal reflects surface properties *(12,13)*. Thus, measurements on nanoscale TIs provide an independent verification of the observed effects in thin films. Previously, a $^{125}$Te NMR study of $Bi_2Te_3$ nanocrystals (see *Main Text*, ref. *7*) as well as $^{209}$Bi NMR spectra of $Bi_2Se_3$ nanowires *(16)* revealed signatures consistent with the TI surface properties of these nanoscale TIs in the limit of higher surface-to-volume ratios. This effect has been discussed in previous NMR studies *(11-13)*, where the reader can find additional discussions of the observed phenomena.

In Figure S6 results from experiments performed on samples of $(Bi_2Te_3)_x(Sb_2Te_3)_{1-x}$ :($Bi_2Te_3$, $Bi_{0.5}Sb_{1.5}Te_3$, $Sb_2Te_3$) and ZnTe are presented as function of average particle size, as estimated from powder x-ray diffraction (PXRD) measurements and the Scherrer formula. The data were acquired with a Bruker DSX-300 spectrometer using a standard Bruker X-nucleus wideline probe with a 5-mm solenoid coil. The [125]Te, [121]Sb and [123]Sb chemical shift scale was calibrated using the unified $\Xi$ scale, relating the nuclear shift to the [1]H resonance of dilute tetramethylsilane in CDCl$_3$ at a frequency of 300.13 MHz. In this study, the [125]Te Knight shift of ball-milled (BM) nanocrystalline Bi$_{0.5}$Sb$_{1.5}$Te$_3$ (TI) was found to be fundamentally different from micrometer size powders obtained by mortar and pestle (m&p). While Bi$_{0.5}$Sb$_{1.5}$Te$_3$ bulk has a regular shift and line profile, nano-Bi$_{0.5}$Sb$_{1.5}$Te$_3$ exhibits a noticeable upfield (negative) Knight shift as function of particle size, which suggests increasing metallic behavior in the limit of large surface-to-volume ratios (Fig. S6A,B). In accordance with the above results, a short $T_1$ and a negative shift have recently been established as the universal NMR features of band inversion in topologically non-trivial materials (TIs) (*15*). In the case of ordinary semiconductors a different behavior has been observed, i.e. the [125]Te frequency shift of BM nano-ZnTe remains equal (un-shifted) to the mortar and pestle sample (Fig. S6C).

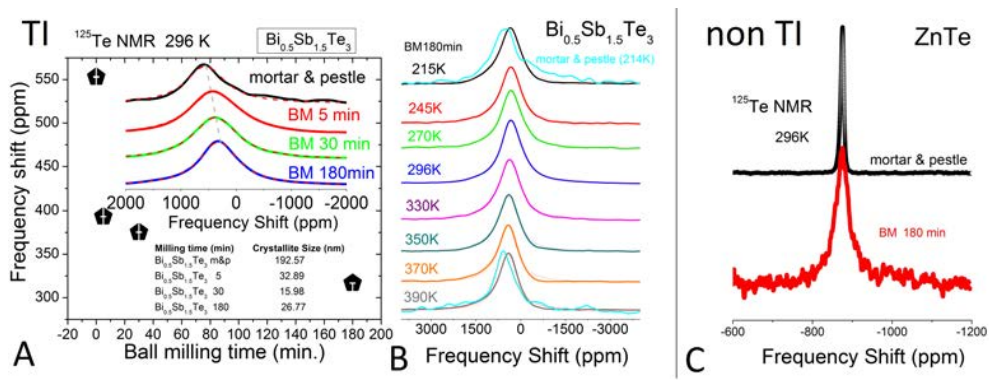

**Fig. S6. NMR study of Bi$_{0.5}$Sb$_{1.5}$Te$_3$ (a TI) and ZnTe (a non TI) materials.** NMR response of the Bi$_{0.5}$Sb$_{1.5}$Te$_3$ reveals negative Knight shift as the crystallite size (**A**) and temperature decreases (**B**) whereas the frequency shift of ZnTe (**C**) remains constant at -875.05 ppm even when the crystallite size is reduced to the nano scale (after a ball milling process lasting 180 min).

The trends in the $\beta$-NMR results are in agreement with the trends from this [125]Te NMR study in Bi$_{0.5}$Sb$_{1.5}$Te$_3$.  The observed 200 ppm frequency shift when going from m&p to the finest ball milled sample is four times larger than the 50 ppm shift observed when varying the beam energy in (Bi,Sb)$_2$Te$_3$ (when going from the TI surface to the TI bulk).  The factor of four could be explained by the different stoichiometry of the two samples and the different nuclear spin probe utilized.  $\beta$-NMR uses quadrupolar nuclei, [8]Li$^+$, which are also sensitive to internal electric field gradients and respond differently to the hyperfine field.

The spin dynamics of quadrupolar nuclei ([121]Sb, [123]Sb) as well as spin-1/2 ([125]Te) of (Bi$_2$Te$_3$)$_x$(Sb$_2$Te$_3$)$_{1-x}$ nanocrystals and bulk materials as function of temperature have also been investigated to shed light in the spin dynamics of the surface states.  The temperature dependence of the [121]Sb and [125]Te spin-lattice relaxation rate for nanoscale (Bi$_2$Te$_3$)$_x$(Sb$_2$Te$_3$)$_{1-x}$ is shown in Fig. S7.  In Fig. S7A the $T_1.T$ product as function of temperature remains constant with temperature in accordance with the Korringa law.  We observed a temperature-independent mechanism indicating the interaction of both [121]Sb and [125]Te spins with conduction carriers over the entire temperature range.  The values of the $T_1.T$ product were 2.7 s.K (for $x$=0) and 11 s.K for ($x$=1).  In Fig. S7B we show a typical relaxation curve of [121]Sb nuclear spin magnetization at 160 K and 7.05 T versus time.  On the other hand, the bulk materials do not follow a Korringa law.  Instead, they show a thermally activated mechanism, which is typical of semiconductors (see Fig. S7C).  The $T_1.T$ product in the bulk is higher than that of nanoscale (Bi$_2$Te$_3$)$_x$(Sb$_2$Te$_3$)$_{1-x}$, as expected: the quantity $1/(T_1.T)$, which is proportional to the square of the density of states at the Fermi level (see *Main Text*, refs. *20-22*), suggests an increased conductivity of the surface compared to the bulk.

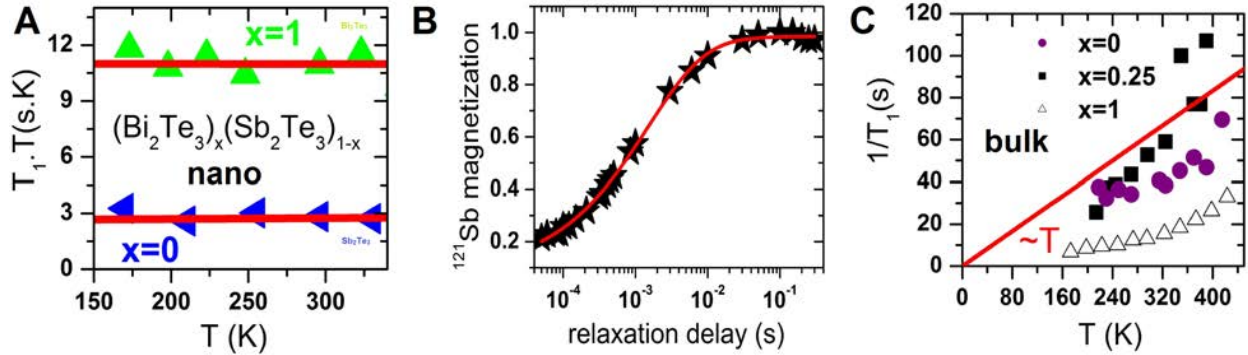

**Fig. S7. NMR $T_1$ and Knight shift study of metallic surface states in $(Bi_2Te_3)_x(Sb_2Te_3)_{1-x}$.** **(A)** $^{121}$Sb and $^{125}$Te NMR spin-lattice relaxation rates ($1/T_1$) of bulk and nanoscale $(Bi_2Te_3)_x(Sb_2Te_3)_{1-x}$ as function of temperature. The Korringa relation holds over the entire temperature range for nanoscale samples of $(Bi_2Te_3)_x(Sb_2Te_3)_{1-x}$, as shown by the red thick lines **(B)** A typical relaxation curve is shown ($^{121}$Sb nuclear spin magnetization at 160 K and 7.05 T versus time). **(C)** In the bulk of these materials, the Korringa law does not hold. Instead, a thermally activated process related to their semiconducting behavior is apparent.

Next, we focus our attention on the second Sb nuclear isotope. $^{123}$Sb has $I=7/2$ and a higher quadrupole moment. We also performed spin-lattice relaxation measurements of $^{123}$Sb. The saturation recovery has been recorded and $T_1$ was determined by fitting the data with a relaxation model for the case of $I=7/2$. The $^{123}$Sb relaxation curves for 255 K, 296 K and 330 K are shown at Fig. S8A-C. We analyzed our data to clarify the underlying mechanism of spin-lattice relaxation of antimony when $x=0$ by calculating the ratios of the relaxation rates of each isotope. In case where the $T_1$ mechanism is governed by fluctuations of the local internal magnetic field at Sb sites from conduction charge carriers, the isotope ratio is proportional to the square of the magnetogyric ratios, as defined in Eq. **S1** below. This magnetic relaxation mechanism is dominant for a ratio of 3.41 (theoretical value).

$$\frac{^{121}(\frac{1}{T_1})}{^{123}(\frac{1}{T_1})} = \frac{^{121}\gamma_n^2}{^{123}\gamma_n^2} = 3.41 \ [\textbf{S1}]$$

Depth-Resolved MRI of TIs

In the case of an electric relaxation mechanism, fluctuations of the electric field gradient (EFG) at the Sb site dominate, and the relaxation rate depends on the nuclear spin $I$. The magnitude of the quadrupolar moment of each Sb isotope plays a dominant role and the ratio is:

$$\frac{^{121}(\frac{1}{T_1})}{^{123}(\frac{1}{T_1})} = \frac{(^{121}Q)^2 \times \frac{(2I+3)}{I^2(2I+1)}}{(^{123}Q)^2 \times \frac{(2I+3)}{I^2(2I+1)}} = 1.44 \ \ [\mathbf{S2}]$$

The experimental ratio of relaxation rates agrees with the theoretical prediction of equation **S1**, namely, fluctuations of the hyperfine field (magnetic) are responsible for the spin-lattice relaxation at the Sb sites as shown in Fig. S8D. Taking into account the Korringa law, this result is further evidence that the interaction of Sb nuclei with the carrier conduction electrons is the dominant mechanism that is responsible for $T_1$ relaxation at the nanoscale for $x$=0 in $(Bi_2Te_3)_x(Sb_2Te_3)_{1-x}$.

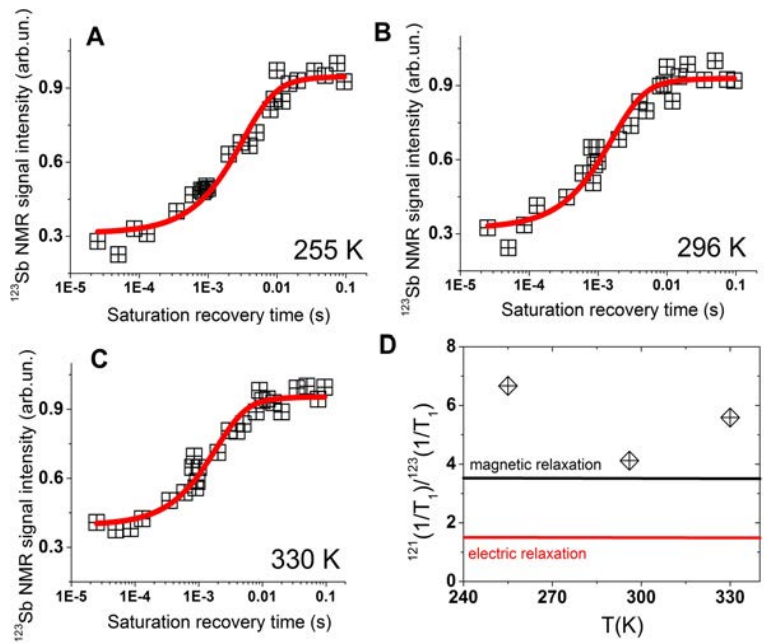

**Fig. S8.** Spin-lattice saturation recovery ($T_1$) relaxation data of $^{123}$Sb at 255 K **(A)**, 296 K **(B)** and 355 K **(C)** for $x$=0. The temperature dependence of the Sb isotope ratio shows that the magnetic interaction is exclusively responsible for the spin-lattice relaxation whereas fluctuations of the electric field gradient at Sb sites do not constitute the dominant mechanism. **(D)**.

Depth-Resolved MRI of TIs

To summarize, $T_1^{-1}$ is governed by the Korringa law over the entire $T$-range. Additionally, the isotope ratio of relaxation rates $^{121}(1/T_1)/^{123}(1/T_1)$ indicates that the nuclear relaxation at Sb sites in the near-surface regime is caused by fluctuations of the hyperfine field (not fluctuations in EFG). These trends are consistent with the observation of metallic states in $(Bi_2Te_3)_x(Sb_2Te_3)_{1-x}$. The observation of metallic behavior in the near-surface regime is consistent with the strategy (demonstrated in transport studies of TIs) that the bulk contribution to conductivity is strongly suppressed in the limit of large surface-to-volume ratios. Although NMR is not a direct probe of the Dirac cone as ARPES is, it does report on the density of states at the Fermi level, which exhibits an important "contrast" mechanism as the bulk contributions from defects are suppressed in the limit of high surface-to-volume ratio.

## D. Magneto-Transport Characterization of $(Bi,Sb)_2Te_3$

The electrical properties of $(Bi,Sb)_2Te_3$ films and their unique topological surface states have been studied by magneto-transport *(1, 19-23)*. In Figure S9A, we observed Shubnikov–de Haas (SdH) oscillations of $-dR_{xx}/dB$ as a function of inverse magnetic field at 0.4 K where the system has entered the quantum regime, and the transport is dominated by the TI surface conduction. To further probe the metallic Dirac surface state, we prepared a 4QL $(Bi_{0.57}Sb_{0.43})_2Te_3$ thin film, and tuned the Fermi level into the surface band gap. When cooled down below 0.3 K, the quantum Hall effect (QHE) is clearly observed in this film: the Hall resistance $R_{xy}$ shows plateaus at the same magnetic fields where the longitudinal resistance $R_{xx}$ develops a minimum (see inset of Fig. S9B). Meanwhile, the Hall resistance $R_{xy}$ shows the expected 2D quantized plateau values, which become clearer in a conductivity plot (see Fig. S9B). As expected for the Dirac surface states where $\sigma_{xy} = 2\,(N+1/2)\,e^2/h$, plateaus of odd filling factor of 5 and 3 can be clearly observed, similar to the quantum Hall effect of an ordinary 2DEG. We note that a recent NMR study on $Bi_2Se_3$ *(17)* has been performed at magnetic fields approaching the quantum limit (i.e. the regime in which only a few of the lowest Landau levels are occupied).

Finally, in light of the unique spin-momentum locking feature of topological surface states, we demonstrate electrical detection of spin-polarized surface state conduction in $(Bi,Sb)_2Te_3$ thin films using a $Co/Al_2O_3$ ferromagnetic tunneling contact, as shown in the inset of Fig. S9C. Specifically, a pronounced resistance hysteresis is observed up to 10 K when sweeping the in-plane magnetic field to change the relative orientation between the Co magnetization and the spin polarization of surface states. The two resistance states are reversible by changing the electric current direction (see Fig. S9D), confirming the spin-momentum locking in the surface state conduction. The spin voltage amplitude was quantitatively analyzed to yield an effective spin polarization of 1.2% for the surface state conduction in the TI. Thus, our TI samples are of the highest available quality and display all of the expected exotic physics of *bona fide* TI materials, as confirmed in this study.

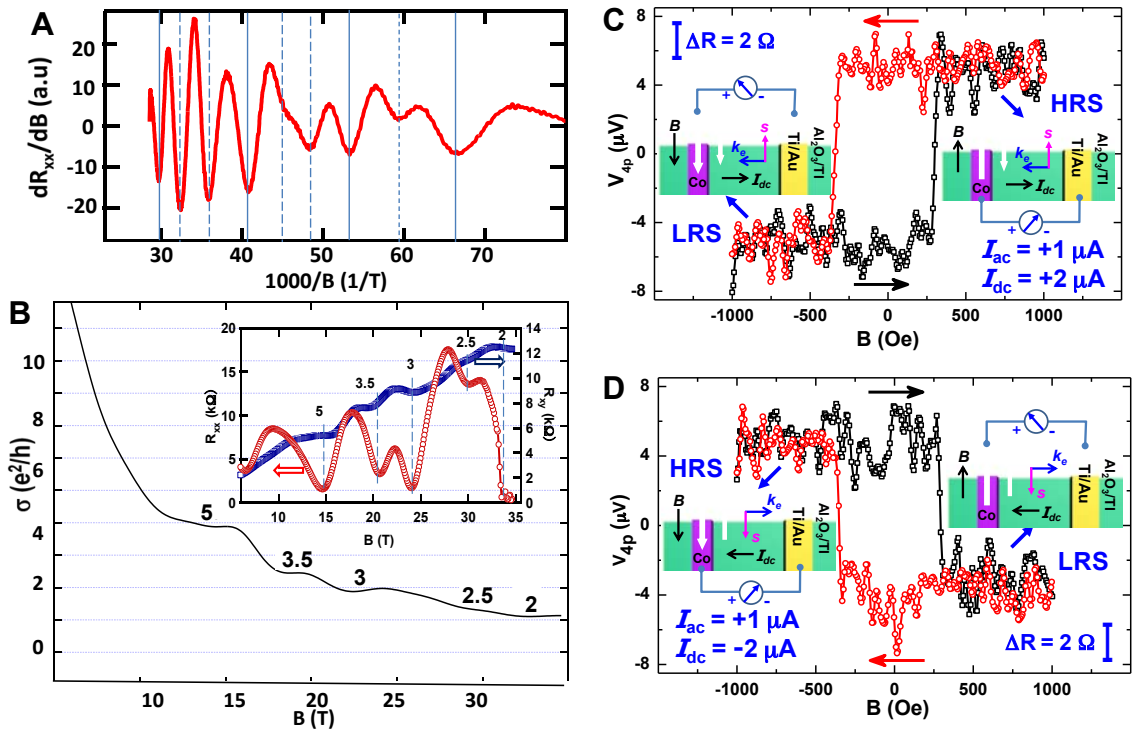

**Fig. S9. Magneto-transport study of (Bi,Sb)$_2$Te$_3$.** **(A)** SdH oscillations of the longitudinal resistance R$_{xx}$ at high magnetic field. **(B)** Quantized Hall effect of the 4QL (Bi$_{0.57}$Sb$_{0.43}$)$_2$Te$_3$ thin film. The Hall conductivity ($\sigma_{xy}$) of the films measured at 0.3 K shows plateaus at the quantized values. The inset shows the Hall resistance R$_{xy}$, together with the longitudinal resistance R$_{xx}$. **(C-D)** Electrical detection of the spin-polarized surface state conduction in (Bi,Sb)$_2$Te$_3$ samples. The measured 4-probe resistance as the in-plane magnetic field is swept back and forth under DC bias of I$_{dc}$ = +2 μA and I$_{dc}$ = -2 μA, respectively. The red and black arrows indicate the magnetic field sweeping direction. The insets show the high-resistance state (HRS) and low-resistance state (LRS), determined by the relative orientation between the Co magnetization and the spin polarization of surface states. See Ref. *(23)*.

**SUPPLEMENTARY REFERENCES**